\documentclass[
aps,
prb,
superscriptaddress,
twocolumn,
10pt,
nofootinbib,
floatfix,
longbibliography
]{revtex4-2}

\usepackage{amssymb}
\usepackage{amsmath}
\usepackage{appendix}
\usepackage[T1]{fontenc}
\usepackage{newtxtext,newtxmath}
\usepackage{microtype}
\usepackage{graphicx}
\usepackage{subeqnarray}
\usepackage{bbold}
\usepackage{enumerate}
\usepackage{color}
\usepackage{bm}
\usepackage[colorlinks=true, letterpaper=true, pdfstartview=FitV, linkcolor=blue, citecolor=blue, urlcolor=blue]{hyperref}
\usepackage[normalem]{ulem}

\newcommand{\backmattersection}[1]{%
  \par\addvspace{1.0\baselineskip}
  \noindent\textbf{#1}\par
  \nobreak\vspace{0.35\baselineskip}
}

\setcitestyle{super,open={},close={}}

\makeatletter
\renewcommand{\@biblabel}[1]{#1.}

\makeatother
\begin{document}
\title{Ferroelectric Band Twinning from Pair-State Symmetry}

\author{Shibo Fang}
\affiliation{Science, Mathematics and Technology (SMT) Cluster, Singapore University of Technology and Design, Singapore 487372}

\author{Jianhua Wang}
\affiliation{Science, Mathematics and Technology (SMT) Cluster, Singapore University of Technology and Design, Singapore 487372}
\affiliation{Institute for Superconducting and Electronic Materials, Faculty of Engineering and Information Sciences, University of Wollongong, Wollongong 2500, Australia}
\affiliation{School of Materials Science and Engineering, Tiangong University, Tianjin 300387, China}

\author{Zhenzhou Guo}
\affiliation{Institute for Superconducting and Electronic Materials, Faculty of Engineering and Information Sciences, University of Wollongong, Wollongong 2500, Australia}

\author{Jialin Gong}
\affiliation{Institute for Superconducting and Electronic Materials, Faculty of Engineering and Information Sciences, University of Wollongong, Wollongong 2500, Australia}

\author{Haiyu Meng}
\affiliation{School of Physics and Optoelectronics, Xiangtan University, Xiangtan 411105, China}

\author{Ruixiang Fei}
\affiliation{Key Laboratory of Advanced Optoelectronic Quantum Architecture and Measurement (MOE), School of Physics, Beijing Institute of Technology, Beijing 100081, China}

\author{Wenhong Wang}
\affiliation{School of Materials Science and Engineering, Tiangong University, Tianjin 300387, China}

\author{Xiaotian Wang}
\email{xiaotianw@uow.edu.au}
\affiliation{Institute for Superconducting and Electronic Materials, Faculty of Engineering and Information Sciences, University of Wollongong, Wollongong 2500, Australia}

\author{Zhenxiang Cheng}
\email{cheng@uow.edu.au}
\affiliation{Institute for Superconducting and Electronic Materials, Faculty of Engineering and Information Sciences, University of Wollongong, Wollongong 2500, Australia}

\author{Yee Sin Ang}
\email{yeesin\_ang@sutd.edu.sg}
\affiliation{Science, Mathematics and Technology (SMT) Cluster, Singapore University of Technology and Design, Singapore 487372}

\begin{abstract}
Ferroelectric switching provides a nonvolatile way to control electronic
structures, but a general symmetry rule connecting the full Bloch bands of
two switchable polarization states is still lacking. Here, we introduce
ferroelectric band twinning, a pair-state relation in which the bands of two
opposite-polarization states are mapped onto each other by a non-inversion
state-exchange symmetry. Using dichromatic groups, we derive the band-twinning rule and identify 11
ferroelectric band-twinning point-group classes. Screening the Ferroelectric Materials Database yields 16 candidate compounds,
of which the two lattice-metric-preserving candidates, bulk
$\gamma$-Ag$_3$SI and BaAl$_2$O$_4$, are selected for first-principles
validation. For $\gamma$-Ag$_3$SI, we further show that the same
pair-state symmetry controls the transformation of shift-current tensor
components under polarization reversal. These results establish ferroelectric band twinning as a general symmetry framework for nonvolatile control of momentum-dependent electronic structures in ferroelectrics.
\end{abstract}

\maketitle
\section{Introduction}
Ferroelectricity enables the nonvolatile reconfiguration of electronic structures through electrically switchable polarization states~\cite{martin2017thin,kim2023wurtzite,scott2007applications}. Recent studies have revealed pronounced polarization-driven band reconstruction in multiferroics, such as altermagnetic multiferroics~\cite{duan2025antiferroelectric,FEAMLiu2025,zhu2025sliding,sun2026altermagnetic,sun2026unified,sun2026sixstate} and other emerging multiferroic states~\cite{eerenstein2006multiferroic,guo2026sliding,song2025electrical,bennett2024stacking,huang2024manipulating}, where polarization reversal reorganizes the spin-split electronic bands. Polarization reversal has also been widely explored for controlling interface effects~\cite{skidmore2025proximity}, spin-to-charge conversion~\cite{varotto2021room}, sliding ferroelectrics~\cite{viznerstern2021interfacial,li2026highly,yasuda2024ultrafast,gaolingyuan2025}, moir\'e ferroelectricity~\cite{kang2023switchable,zheng2020unconventional}, bulk photovoltaic effects~\cite{feng2025high,choi2009switchable}, and  nonvolatile memories~\cite{quhe2024asymmetric,jia2024giant,tsymbal2006tunneling}. Yet the most elementary case---a nonmagnetic ferroelectric whose two states
differ only in the sign of $\mathbf{P}$---has rarely been examined as a general rule: how the two Bloch-band structures are related is usually
addressed case by case for a specific material or a specific band feature~\cite{li2026highly,kang2023switchable,quhe2024asymmetric}.

Conventional crystallographic descriptions are usually formulated for a single state~\cite{michel1980symmetry}: An individual ferroelectric
state is typically characterized by the crystallographic space group of its
polar structure, or by the corresponding crystallographic point group when
only macroscopic properties are considered
~\cite{junquera2023topological,ji2023general}. Ferroelectric switching, by
contrast, is intrinsically relational: connecting selected initial and final polarization states along a continuous path
~\cite{liang2025resolving,zhang2023ferroelectric,gou2023two,fei2018ferroelectric,wu2023unconventional}. In the modern
theory of polarization, the absolute polarization of a periodic crystal is
defined only modulo a polarization quantum, whereas the polarization change
along such a path is well defined~\cite{Resta1994,KingSmith1993}. Recent
discussions of a unified definition of ferroelectricity have also placed
the emphasis on an electrically switchable pair state rather than an isolated
polar structure~\cite{luo2026unified}. These developments motivate a symmetry description of pairs of switchable ferroelectric states.

\begin{figure*}[t]
  \centering
  \includegraphics[width=\textwidth]{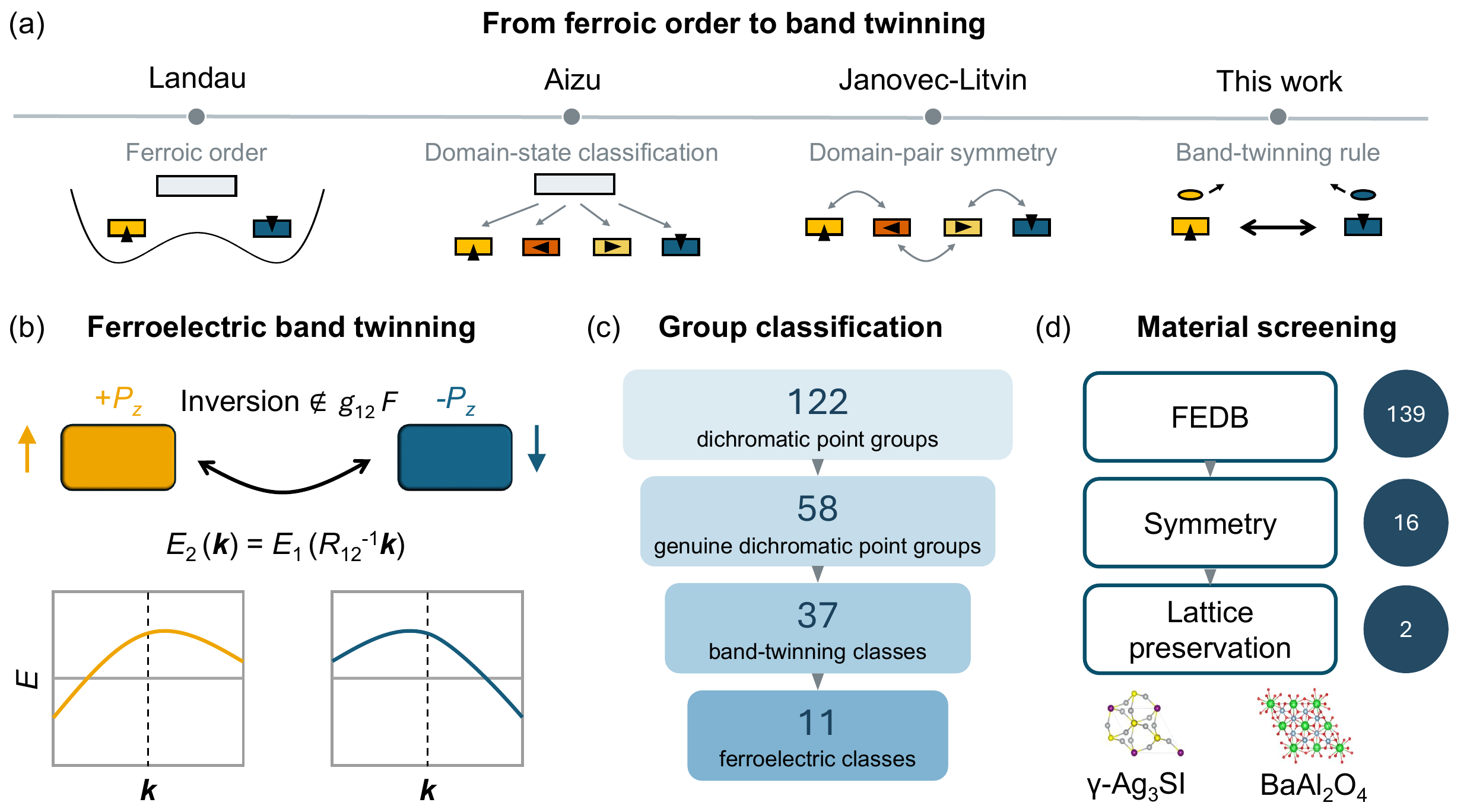}
  \caption{\textbf{Concept and workflow of ferroelectric band twinning.}
\textbf{a}, Relationship between ferroic order, domain-state classification, domain-pair symmetry, and band twinning.
\textbf{b}, Schematic of two switchable polarization states and their symmetry-related band structures. Here,
\(g_{12}=\{R_{12}|\boldsymbol{\tau}\}\) denotes the state-exchange
operation, with \(R_{12}\) acting on crystal momentum.
\textbf{c}, Symmetry classification of ferroelectric band-twinning point-group classes.
\textbf{d}, Symmetry-based screening of candidate ferroelectric materials.}
  \label{fig1}
\end{figure*}

Early ferroic domain theory already provides the conceptual basis for a
pair-state description. Within the Landau-Aizu framework, ferroic domain
states arise through symmetry lowering from a higher-symmetry parent phase
and are classified by the corresponding cosets
~\cite{pokrovsky2009landau,aizu1969possible,Aizu1970}. Janovec subsequently
extended this single-domain description to symmetry-distinct relations
between pairs of ferroic states using double-coset decomposition
~\cite{janovec1972group,Janovec1976}. For a selected pair state, Janovec,
Litvin, and co-workers introduced a dichromatic group whose operations either
preserve each state individually or exchange the two states. These frameworks, however, were developed to describe real-space domain relations and macroscopic property tensors~\cite{janovec1994tensor}. Their implications for Bloch Hamiltonians and momentum-dependent electronic structures have not been systematically developed.

Here we extend the symmetry description of ferroic pair states from real-space domain relations to momentum-space electronic bands. We introduce ferroelectric band twinning as a pair-state relation in which the Bloch-band dispersions of two switchable polarization states are mapped onto each other by a non-inversion state-exchange operation, so that in a common crystallographic and momentum frame the two dispersions are symmetry-related but generally differ at the same crystal momentum. By formulating the action of
the state-exchanging coset of a dichromatic group on the corresponding Bloch
Hamiltonians, we derive the band-twinning rule and identify 11 ferroelectric
band-twinning point-group classes. Symmetry-guided database screening yields
16 candidate compounds, among which bulk $\gamma$-Ag$_3$SI and
BaAl$_2$O$_4$ are the only two for which lattice-metric preservation is enforced by pair-state symmetry. First-principles calculations confirm the predicted
band reconfiguration in both materials, and further show that the same
pair-state symmetry governs the transformation of shift-current tensor
components under polarization reversal. 

\section{Results}
\subsection{Pair-state symmetry and the theory of ferroelectric band twinning}
The pair-state symmetry framework underlying band twinning is summarized in Fig.~\ref{fig1}(a): Aizu's coset decomposition organizes the ferroic domain states generated from a common parent phase~\cite{aizu1969possible,Aizu1970}, Janovec's double-coset decomposition classifies the symmetry-distinct relations between pairs of such states~\cite{janovec1972group,Janovec1976}, and the Janovec–Litvin dichromatic framework separates, for a selected pair state, the operations that preserve each state from those that exchange the two states~\cite{janovec1994tensor}. Here, we extend this pair-state framework from real-space domain relations and macroscopic property tensors to pairs of Bloch Hamiltonians, leading to the band-twinning rule in momentum space; further details and a comparison of these symmetry frameworks are provided in the Supplementary Information.

\begin{figure*}[t]
\centering
\includegraphics[width=\textwidth]{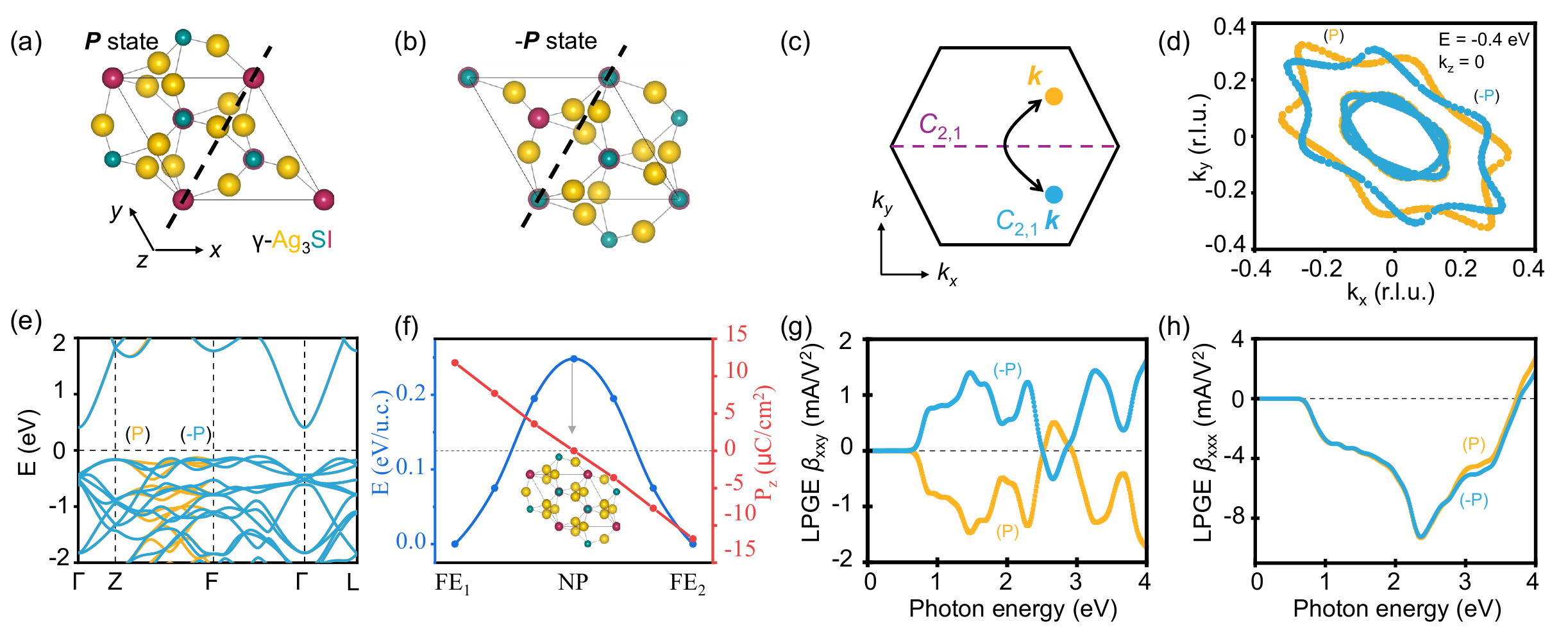}
\caption{
\textbf{Ferroelectric band twinning in \(\gamma\)-Ag$_3$SI.}
\textbf{a,b}, Crystal structures of the \(\mathbf{P}\) and \(-\mathbf{P}\) states.
\textbf{c}, Momentum-space mapping under \(C_{2,1}\).
\textbf{d}, Constant-energy contours at \(E=-0.4\) eV and \(k_z=0\). r.l.u. denotes reciprocal lattice units.
\textbf{e}, Electronic band structures of the two states.
\textbf{f}, Energy and polarization evolution along the considered symmetry-compatible switching pathway.
\textbf{g,h}, Shift-current spectra of the linear photogalvanic effect (LPGE) coefficients \(\beta_{xxy}\) and \(\beta_{xxx}\), respectively.
}
\label{fig2}
\end{figure*}

We define band twinning as a switching-induced relation between two distinct physical states of the same system that can be interconverted through processes such as ferroelectric reversal or ferroelastic switching. Described in a common crystallographic and momentum frame, the two
states are exchanged by an operation
\(g_{12}=\{R_{12}|\boldsymbol{\tau}\}\notin F\), whereas the operations
in \(F\) preserve each state individually. Here, \(R_{12}\) denotes the
point-group part of \(g_{12}\), which acts on crystal momentum, while
the translational part \(\boldsymbol{\tau}\) is incorporated into the
\(\mathbf{k}\)-dependent representation \(U_{12}(\mathbf{k})\). Their
Bloch Hamiltonians therefore satisfy
\(H_2(\mathbf{k})=U_{12}(\mathbf{k})H_1(R_{12}^{-1}\mathbf{k})
U_{12}^{-1}(\mathbf{k})\), yielding
\(E_{2,n}(\mathbf{k})=E_{1,n}(R_{12}^{-1}\mathbf{k})\): the two band structures are exactly related by the symmetry-prescribed momentum mapping but generally non-identical at the same \(\mathbf{k}\). This relation represents an active transformation
between two switchable physical states, rather than a passive
coordinate or basis transformation of a single state.

\begin{table}[b]
\centering
\caption{
The 11 ferroelectric band-twinning point-group classes.
Primed operations exchange the two ferroelectric states.
}
\label{tab:fe-band-twinning-classes}

\small
\renewcommand{\arraystretch}{1.12}
\setlength{\tabcolsep}{0pt}

\begin{tabular*}{\columnwidth}{@{\extracolsep{\fill}}lc@{}}
\hline
Crystal system
&
Ferroelectric band-twinning point-group classes
\\
\hline

Monoclinic
&
\(2^{\prime},\quad m^{\prime}\)
\\

Orthorhombic
&
\(2^{\prime}2^{\prime}2,\quad 2^{\prime}m^{\prime}m\)
\\

Tetragonal
&
\(\bar{4}^{\prime},\quad
42^{\prime}2^{\prime\dagger},\quad
\bar{4}^{\prime}m2^{\prime}\)
\\

Trigonal
&
\(32^{\prime\dagger}\)
\\

Hexagonal
&
\(\bar{6}^{\prime\dagger},\quad
62^{\prime}2^{\prime\dagger},\quad
\bar{6}^{\prime}m2^{\prime\dagger}\)
\\

\hline
\end{tabular*}

\vspace{3pt}

\parbox{\columnwidth}{%
\footnotesize
\raggedright
\(\dagger\) Preservation of the lattice metric is enforced by
pair-state symmetry. For the remaining classes, identical lattice
metrics are not guaranteed by point-group symmetry alone.
}

\end{table}

For the time-reversal-symmetric pairs considered here, time-reversal symmetry imposes $E_n(\mathbf{k})=E_n(-\mathbf{k})$. If the exchange operation is inversion, then $E_{2,n}(\mathbf{k})=E_{1,n}(-\mathbf{k})=E_{1,n}(\mathbf{k})$, and switching leaves the band dispersion unchanged. By contrast, a non-inversion exchange operation generally maps $\mathbf{k}$ to a symmetry-distinct momentum, such that $E_{2,n}(\mathbf{k})=E_{1,n}(R_{12}^{-1}\mathbf{k})\neq E_{1,n}(\mathbf{k})$ at fixed momentum coordinates. For the lattice-preserving switching processes considered here, the Bravais lattice and momentum-coordinate frame remain unchanged, so the switching-induced band reconfiguration can be compared directly between the two states.

We next classify which state-pair symmetries can support band twinning. The
122 dichromatic point groups comprise 32 ordinary, 32 gray, and 58 black-white
groups. Ordinary groups contain no state-exchanging operations, whereas gray
groups contain the spatial identity combined with state exchange and therefore
do not describe a nontrivial spatial relation between two distinct states.
Only the 58 black-white groups, corresponding to the genuine dichromatic
point groups, admit the disjoint decomposition
\(M=F+g_{12}F\), where \(F\) preserves each state and \(g_{12}F\) exchanges two states. Band twinning further requires that the state-exchanging
coset \(g_{12}F\) exclude inversion: if inversion exchanges the two states, time-reversal symmetry makes their band dispersions identical at the same
\(\mathbf{k}\). Excluding the 21 genuine
dichromatic point groups whose state-exchanging cosets contain inversion
leaves 37 symmetry-allowed classes, which we define as the
\emph{band-twinning point-group classes}.

For a ferroelectric pair state, band twinning additionally requires two
physically switchable states with opposite polarizations,
\(\mathbf{P}\rightarrow-\mathbf{P}\). For a conventional ferroelectric
pair, the state-preserving subgroup \(F\) must admit a nonzero
polarization that is invariant under every operation in \(F\), whereas
the state-exchanging coset must reverse this polarization. These
conditions can be written as
\begin{equation}
R_f\mathbf{P}=\mathbf{P}
\quad (f\in F),
\qquad
R_{g_{12}f}\mathbf{P}=-\mathbf{P}
\quad (g_{12}f\in g_{12}F).
\label{eq:ferroelectric-pair-criterion}
\end{equation}
Because every element of the state-exchanging coset has the form
\(g_{12}f\), the second relation follows from
\(R_{g_{12}}\mathbf{P}=-\mathbf{P}\) and
\(R_f\mathbf{P}=\mathbf{P}\). 

The classification of ferroelectric band twinning proceeds through three
symmetry-based filters. Starting from the 37 general band-twinning classes,
we first exclude 12 classes whose state-exchanging cosets contain a proper
rotation \(C_n\) of order \(n>2\), because such a rotation cannot map any
nonzero polar vector directly onto its negative. This leaves 25 candidate
classes. We then exclude 10 classes whose state-preserving subgroups \(F\)
are nonpolar, reducing the number to 15. Finally, four classes,
\(m^\prime m^\prime 2\), \(4m^\prime m^\prime\),
\(6m^\prime m^\prime\), and \(3m^\prime\), are excluded because their
state-exchanging mirrors preserve, rather than reverse, the polarization
allowed by \(F\). The remaining 11 classes constitute the
\emph{ferroelectric band-twinning point-group classes}, as summarized in
Table~\ref{tab:fe-band-twinning-classes}. Further details of the filtering
procedure are provided in the Supplementary Information.

The above 11-class classification is based on the polarization-reversal
condition alone. Inspired by the concept of ferroelastic twins~\cite{janovec1994tensor}, the classification can be further refined by considering whether the lattice metric is preserved.
Since state 2 is obtained from state 1 through a state-exchange operation,
their spontaneous strain tensors satisfy
\(\boldsymbol{\varepsilon}_2=
R_{12}\boldsymbol{\varepsilon}_1R_{12}^{T}\).
Identical relaxed lattice metrics are guaranteed by symmetry only when every
strain tensor allowed by the state-preserving subgroup \(F\) is individually
invariant under the state-exchange operations, namely
\(R_{12}\boldsymbol{\varepsilon}R_{12}^{T}
=\boldsymbol{\varepsilon}\).
This condition is satisfied in the five classes
\(42^{\prime}2^{\prime}\), \(32^{\prime}\),
\(\bar{6}^{\prime}\), \(62^{\prime}2^{\prime}\), and
\(\bar{6}^{\prime}m2^{\prime}\).
In the remaining six classes,
\(2^{\prime}\), \(m^{\prime}\), \(2^{\prime}2^{\prime}2\),
\(2^{\prime}m^{\prime}m\), \(\bar{4}^{\prime}\), and
\(\bar{4}^{\prime}m2^{\prime}\), the subgroup \(F\) permits anisotropic
or shear strains that may be reversed or interchanged by the state-exchange
operation, so identical relaxed lattice metrics are not guaranteed by
point-group symmetry alone. Nevertheless, the band-twinning relation
\(E_{2,n}(\mathbf{k})=
E_{1,n}(R_{12}^{-1}\mathbf{k})\) remains exact between the corresponding
symmetry-related Brillouin zones. A point-by-point comparison at the same
\(\mathbf{k}\) in a common Brillouin zone additionally requires lattice
clamping or negligible spontaneous-strain differences.

\subsection{Symmetry-guided discovery and material realizations}
We now apply the symmetry classification to identify band-twinning ferroelectric pair states, defined here as pairs of switchable ferroelectric states that obey the band-twinning relation. First, the point group of a polar ferroelectric state is identified with the state-preserving subgroup \(F\) of the 11 ferroelectric
band-twinning point-group classes. Materials satisfying this subgroup condition are
then examined at the space-group level to identify possible
state-exchanging operations. Second, a symmetry-related partner state is
constructed by applying an operation from the state-exchanging coset
\(g_{12}F\), whose elements exchange the two states. Third, the resulting symmetry-allowed candidates are validated by first-principles calculations of their switching pathways, polarizations, and switching-induced band reconfigurations.

We implement this procedure using the Ferroelectric Materials Database
(FEDB)~\cite{Smidt2020}. Among 413 switching-pathway entries, we retain 200 entries with completed calculations and smooth switching profiles, corresponding to 139 distinct compounds. Examining the state-preserving point group \(F\) of the polar state and the
corresponding state-exchanging coset \(g_{12}F\) for each entry identifies
16 symmetry-allowed candidate compounds among the 11 ferroelectric
band-twinning point-group classes, as summarized in Table~\ref{tab:fe-band-twinning-candidates}. Several compounds admit more
than one symmetry-compatible switching branch, with the corresponding
pair states belonging to different ferroelectric band-twinning
point-group classes. For first-principles validation, we prioritize
candidates for which preservation of the lattice metric is enforced by
pair-state symmetry, allowing the two polarization states to be compared
in a common momentum frame without invoking lattice clamping. Among the
16 FEDB candidates, $\gamma$-Ag$_3$SI in the \(32^{\prime}\) class and
BaAl$_2$O$_4$ in the \(62^{\prime}2^{\prime}\) class are the only
candidates satisfying this additional symmetry criterion and are therefore
selected as representative band-twinning ferroelectrics.

\begin{table}[b]
\centering
\caption{
The 16 symmetry-allowed conventional ferroelectric band-twinning
candidates identified from the completed calculations and smooth switching profiles in the FEDB~\cite{Smidt2020}. The pair-state class associated with
each retained switching branch is listed below the material.
H$_2$O denotes the polar ice.
}
\label{tab:fe-band-twinning-candidates}

\footnotesize
\setlength{\tabcolsep}{1pt}

\newcommand{\matcell}[2]{%
\parbox[c][2.8em][c]{0.235\columnwidth}{%
\centering
\strut #1\\[-1pt]
\strut #2
}}

\begin{tabular}{@{}cccc@{}}
\hline

\matcell{$\gamma$-Ag$_3$SI}
        {$32^{\prime}$}
&
\matcell{BaAl$_2$O$_4$}
        {$62^{\prime}2^{\prime}$}
&
\matcell{CuI}
        {$2^{\prime}m^{\prime}m$}
&
\matcell{GaN$_5$O$_{14}$}
        {$\bar{4}^{\prime}$}
\\[2pt]

\matcell{H$_2$O}
        {$2^{\prime}2^{\prime}2$}
&
\matcell{K$_3$SbS$_4$}
        {$\bar{4}^{\prime}m2^{\prime}$}
&
\matcell{K$_4$CO$_4$}
        {$\bar{4}^{\prime},\;2^{\prime}m^{\prime}m$}
&
\matcell{KH$_2$PO$_4$}
        {$\bar{4}^{\prime}m2^{\prime}$}
\\[2pt]

\matcell{Li$_4$CO$_4$}
        {$\bar{4}^{\prime}$}
&
\matcell{Li$_6$PS$_5$I}
        {$2^{\prime}m^{\prime}m$}
&
\matcell{Mg$_3$B$_7$O$_{13}$Cl}
        {$\bar{4}^{\prime}m2^{\prime}$}
&
\matcell{Na$_4$CO$_4$}
        {$\bar{4}^{\prime},\;2^{\prime}m^{\prime}m$}
\\[2pt]

\matcell{Rb$_4$CO$_4$}
        {$2^{\prime}2^{\prime}2,\;\bar{4}^{\prime}$}
&
\matcell{RbH$_2$PO$_4$}
        {$\bar{4}^{\prime}m2^{\prime}$}
&
\matcell{SiO$_2$}
        {$2^{\prime}2^{\prime}2,\;\bar{4}^{\prime}m2^{\prime}$}
&
\matcell{Sn$_3$O$_2$(OH)$_2$}
        {$2^{\prime}m^{\prime}m$}
\\

\hline
\end{tabular}
\end{table}

\begin{figure}[t]
\centering
\includegraphics[width=\columnwidth]{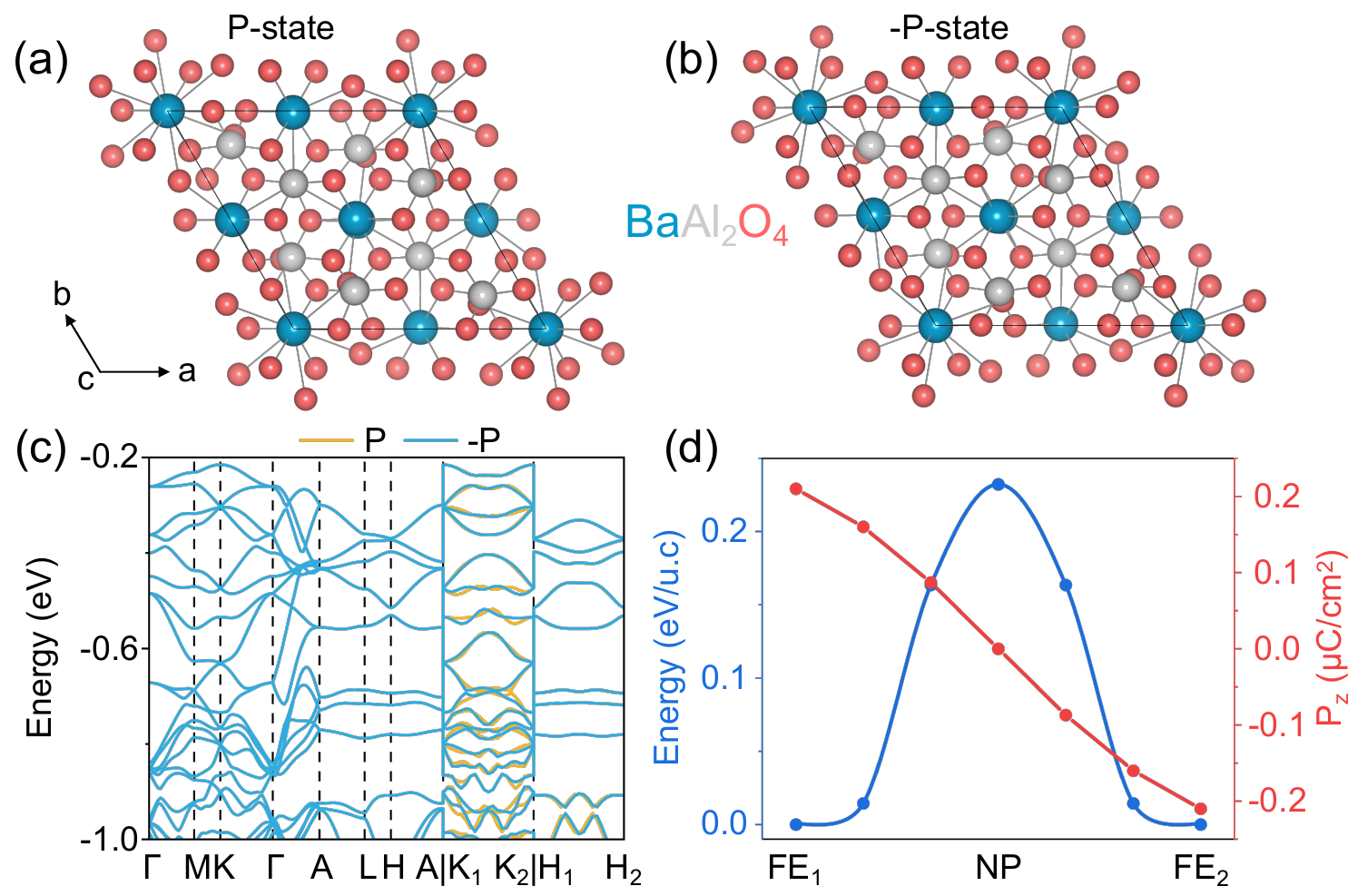}
\caption{
\textbf{Ferroelectric band twinning in BaAl$_2$O$_4$.}
\textbf{a,b}, Crystal structures of the $\mathbf{P}$ and
$-\mathbf{P}$ states.
\textbf{c}, Electronic band structures of the two polarization states,
showing the switching-induced band reconfiguration.
\textbf{d}, Energy profile and polarization component $P_z$ along the
FE$_1$-NP-FE$_2$ switching pathway.
}
\label{fig3}
\end{figure}

Ag$_3$SI is an experimentally synthesized chalcohalide antiperovskite whose
low-temperature $\gamma$ phase adopts the polar \(R3\) structure~\cite{Hoshino1979,Hull2001}, for which the switchable ferroelectricity was subsequently predicted~\cite{Ma2024}. As shown in
Fig.~\ref{fig2}(a,b), the two ferroelectric states carry opposite
polarizations along the crystallographic \(c\) axis and are related by a
twofold rotation about an axis in the basal plane. In the pair-state
description, this pair state belongs to the dichromatic point group
\(32^\prime\), and a representative of the state-exchanging coset can be
chosen as \(g_{12}=C_{2,1}\) [Fig.~\ref{fig2}(c)]. Because
\(C_{2,1}\) reverses the polarization,
\(C_{2,1}\mathbf{P}=-\mathbf{P}\), and is not a spatial inversion
operation, it satisfies the state-exchange condition for ferroelectric
band twinning, and two states obey
\(E_2(\mathbf{k})=E_1(C_{2,1}^{-1}\mathbf{k})\).

First-principles calculations confirm the symmetry-prescribed band
twinning. At \(E=-0.4\) eV in the \(k_z=0\) plane, the
constant-energy contours of the \(\mathbf{P}\) and \(-\mathbf{P}\)
states are mapped onto each other by \(C_{2,1}\), rather than
coinciding at fixed momentum coordinates [Fig.~\ref{fig2}(d)]. The calculated bands exhibit the same
mapping [Fig.~\ref{fig2}(e)]: the spectra coincide when
\(C_{2,1}^{-1}\mathbf{k}=\mathbf{k}\) or
\(C_{2,1}^{-1}\mathbf{k}=-\mathbf{k}\), the latter following from
time-reversal symmetry, whereas at generic momenta they are interchanged
between \(C_{2,1}\)-related paths. The two polar states are connected by
the considered symmetry-compatible pathway through an intermediate
nonpolar configuration [Fig.~\ref{fig2}(f)], along which
\(P_z\) reverses between
\(\pm 11.8~\mu\mathrm{C/cm^2}\) with an energy barrier of
\(0.25~\mathrm{eV}\) per unit cell. 

Such pair-state symmetry also constrains the shift-current
response~\cite{Dai2023,YangLiuFEmetal,young2012first}. The bulk photovoltaic shift current is
described by the third-rank polar tensor \(\beta_{abc}\), whose
components in the two ferroelectric states are related by the
state-exchange operation when compared in the same fixed coordinate
frame. For \(\gamma\)-Ag$_3$SI, \(C_{2,1}\) acts as
\((x,y,z)\rightarrow(x,-y,-z)\). Each \(y\) or \(z\) index therefore
contributes a sign change, so a component is preserved when the total
number of \(y\) and \(z\) indices is even and reverses sign when it is
odd. Accordingly, \(\beta_{xxy}\) reverses sign between the two
ferroelectric states [Fig.~\ref{fig2}(g)], whereas \(\beta_{xxx}\)
remains nearly invariant [Fig.~\ref{fig2}(h)]. By
contrast, for a conventional ferroelectric pair related by inversion,
all components of the third-rank polar tensor reverse sign upon
polarization reversal. The persistence of a switching-even component
such as \(\beta_{xxx}\) therefore provides a possible symmetry-resolved
signature of ferroelectric band twinning. In principle, this signature
could be examined in the low-temperature \(R3\) phase of
\(\gamma\)-Ag$_3$SI by comparing the polarization-dependent
photocurrents of oppositely polarized domains using spatially resolved
measurements, without requiring complete macroscopic poling~\cite{lee2012spatially,chang2023shift}. A possible electrode geometry, polarization protocol, and
domain-resolved signal-extraction scheme are described in Supplementary.

As a complementary intrinsic example, we consider BaAl$_2$O$_4$, an experimentally reported ferroelectric oxide whose polar \(P6_3\)
phase emerges from the nonpolar \(P6_322\) parent
structure~\cite{Huang1994BaAl2O4,Huang1994Phase}. Its two
opposite-polarization states form the dichromatic point-group class
\(62^{\prime}2^{\prime}\): the sixfold rotation about the polar
crystallographic \(c\) axis preserves each state, whereas the twofold
rotations about axes in the basal plane exchange them. Accordingly, the
degenerate \(\mathbf{P}\) and \(-\mathbf{P}\) states carry opposite
polarizations along the \(c\) axis and are related by a twofold rotation [Fig.~\ref{fig3}(a,b)]. 

First-principles calculations show that their electronic structures obey the corresponding band-twinning relation [Fig.~\ref{fig3}(c)]: although the two states generally have different band energies at the same
momentum, their bands are mapped onto each other by the state-exchange
operation, which interchanges the Brillouin-zone corner momenta \(K_1\) and
\(K_2=C_2K_1\). The calculated switching pathway connects the two degenerate
polar states through an intermediate nonpolar configuration, denoted
NP [Fig.~\ref{fig3}(d)]. Along this pathway, \(P_z\) reverses from
approximately \(+0.22\) to \(-0.22~\mu\mathrm{C/cm^2}\), with an energy
barrier of \(0.24~\mathrm{eV}\) per unit cell near the nonpolar
configuration. 

In summary, we have established ferroelectric band twinning as a
symmetry-based relation between the Bloch bands of two switchable states in
time-reversal-symmetric ferroelectrics. Extending dichromatic-group symmetry
from real-space ferroic pair states to pairs of Bloch Hamiltonians yields the
band-twinning rule and 11 ferroelectric band-twinning point-group classes,
which are validated by first-principles calculations in the intrinsic bulk
ferroelectrics $\gamma$-Ag$_3$SI and BaAl$_2$O$_4$. Our work provides a
symmetry framework for the nonvolatile reconfiguration of band structures in
switchable ferroelectrics.

\section{Methods}
First-principles calculations were performed within density functional theory
using the PBE generalized-gradient approximation and the PAW method, as
implemented in VASP~\cite{Kresse1999,Kresse1996a,Kresse1996b}, with a
plane-wave cutoff of 500~eV and an electronic-energy convergence threshold
of \(10^{-6}\)~eV for both structural relaxations and final static
calculations. Structural relaxations were continued until the residual
forces on each atom fell below
\(0.01~\mathrm{eV}\,\text{\AA}^{-1}\). A \(9\times9\times6\)
Monkhorst-Pack mesh was used for the calculations. Spontaneous
polarizations were evaluated using the Berry-phase
method~\cite{KingSmith1993}, and switching pathways were obtained using the
climbing-image nudged elastic band method with five intermediate
images~\cite{Henkelman2000}. VASPKIT and qvasp were used for post-processing
and analysis~\cite{wang2021vaspkit,yi2020qvasp}.
The electronic band structures shown in the main text were calculated
without spin--orbit coupling. Spin-orbit coupling was included only in the
Wannier-based shift-current calculations for \(\gamma\)-Ag$_3$SI. Maximally
localized Wannier functions were constructed using
Wannier90~\cite{Mostofi2008,Mostofi2014,Pizzi2020}, and the resulting Wannier
Hamiltonians were used to interpolate the electronic bands and calculate
the shift-current conductivity~\cite{ShiftcurrentWannier}. A
\(21\times21\times21\) mesh was used for Brillouin-zone integration, with
convergence verified up to \(51\times51\times51\), as shown in Fig.~S1. The electronic band structures of the two polarization states of Ag$_3$SI calculated with spin–orbit coupling, together with comparisons between the VASP and Wannier-interpolated bands, are presented in Fig.~S2.

\section*{Data availability}
The data that support the findings of this study are available within the
Article and its Supplementary Information. Additional data are available
from the corresponding authors upon reasonable request. The ferroelectric
materials data used for materials screening are publicly available from the
Ferroelectric Materials Database cited in the manuscript.

\bibliography{apssamp}

\raggedbottom

\backmattersection{Acknowledgements}
\noindent
This work is supported by the Singapore Ministry of Education (MOE)
Academic Research (AcRF) Tier 2 grant under the award number
MOE-T2EP50125-0019. Y.S.A. acknowledges the support from the Kwan Im
Thong Hood Cho Temple Early Career Chair Professorship. X.W. thanks the
Australian Research Council Discovery Early Career Researcher Award
(Grant No.~DE240100627) for support. Z.C. acknowledges support from the Australian Research Council under Grant No. DP260102992. R.F. is supported by the National Natural Science Foundation of China under Grant No.~12204035. J.W. thanks the China Scholarship Council (CSC).

\backmattersection{Author contributions}
\noindent
S.F. conceived the project, developed the pair-state symmetry framework,
performed the symmetry analysis and first-principles calculations,
analyzed the results, and wrote the original manuscript. J.W. and Z.G.
performed extensive first-principles calculations, processed the
computational data, and prepared and refined the figures. J.G. contributed
to the first-principles calculations. H.M., R.F., and W.W. contributed to
the discussion and interpretation. X.W., Z.C., and Y.S.A. supervised the
project. S.F., X.W., Z.C., and Y.S.A. revised the manuscript with input
from all authors. All authors discussed the results and reviewed and
approved the submitted version.

\backmattersection{Competing interests}
\noindent
The authors declare no competing interests.

\end{document}